%% file: submit.tex
\documentclass[twocolumn,showpacs,aps,prb,reprint,amsfonts,amsmath,amssymb,floatfix,superscriptaddress, longbibliography]{revtex4-2} 

\usepackage{siunitx}
\usepackage{color}
\usepackage{hhline}
\usepackage{mathrsfs}

\usepackage{graphicx}
\usepackage{adjustbox}
\usepackage{dcolumn}
\usepackage{bm}
\usepackage{multirow}
\usepackage{booktabs}
\usepackage{afterpage}
\usepackage{amsmath}
\usepackage{physics}
\usepackage[normalem]{ulem}
\usepackage{here} 
\usepackage{caption}
\usepackage[subrefformat=parens]{subcaption}
\usepackage[version=4]{mhchem}
\usepackage[nohyperlinks,nolist]{acronym} 
\usepackage{multirow}

\usepackage[colorlinks=true,citecolor=blue,linkcolor=blue,urlcolor=blue]{hyperref}
\usepackage{cleveref}
\crefname{equation}{Eq.}{Eq.}
\crefname{figure}{Fig.}{Fig.}
\crefname{table}{Table}{Table}
\crefname{section}{Sec.}{Sec.}
\crefname{appendix}{Appendix}{Appendix}
\Crefname{equation}{Equation}{Equation}
\Crefname{figure}{Figure}{Figure}
\Crefname{table}{Table}{Table}
\Crefname{section}{Section}{Section}
\Crefname{appendix}{Appendix}{Appendix}

\usepackage{threeparttable}

\begin{document}

\title{A chemical bond-based machine learning model for dipole moment: \\
Application to dielectric properties of liquid methanol and ethanol}

\author{Tomohito Amano}
 \affiliation{Department of Physics, The University of Tokyo, Hongo, Bunkyo-ku, Tokyo 113-0033, Japan}
\author{Tamio Yamazaki}
 \affiliation{JSR-UTokyo Collaboration Hub, CURIE, JSR Corporation, 1-9-2, Higashi-Shinbashi, Minato-ku, Tokyo 105-8640, Japan}
\author{Shinji Tsuneyuki}
 \affiliation{Department of Physics, The University of Tokyo, Hongo, Bunkyo-ku, Tokyo 113-0033, Japan}

\date{\today}
 
\onecolumngrid
\begin{abstract}
We introduce a versatile machine-learning scheme for predicting dipole moments of molecular liquids to study dielectric properties. We attribute the center of mass of Wannier functions, called Wannier centers, to each chemical bond and create neural network models that predict the Wannier centers for each chemical bond. Application to liquid methanol and ethanol shows that our neural network models successfully predict the dipole moment of various liquid configurations in close agreement with DFT calculations. We show that the dipole moment and dielectric constant in the liquids are greatly enhanced by the polarization of Wannier centers due to local intermolecular interactions. The calculated dielectric spectra agree well with experiments quantitatively over terahertz (THz) to infrared regions. Furthermore, we investigate the physical origin of THz absorption spectra of methanol, confirming the importance of translational and librational motions. Our method is applicable to other molecular liquids and can be widely used to study their dielectric properties.
\end{abstract}

\twocolumngrid

\maketitle

\begin{acronym}[PPG725]
\setlength{\itemsep}{4.0pt}
     \acro  {H-bond}[H-bond] {hydrogen bond} 
     \acrodefplural{H-bond}[H-bonds] {hydrogen bonds} \acro  {RMSE}  [RMSE]   {root mean square error}
     \acro  {RDF}   [RDF]    {radial distribution function}
     \acro  {KS}    [KS]    {Kohn-Sham} \acro  {O-lp}  [O-lp]  {O lone-pair} \acro  {DFPT}  [DFPT]  {density functional perturbation theory} \acro  {DFT}   [DFT]   {density functional theory}
     \acro  {ML}    [ML]    {machine learning}
     \acro  {VACF}  [VACF]  {velocity auto-correlation function}
     \acro  {PG}    [PG]    {propylene glycol }
     \acro  {PG2}   [PG2]   {di-propylene glycol }
     \acro  {PPG725}[PPG725]{14-mer of Propylene glycol }
     \acro  {SCPH}  [SCPH]  {self-consistent phonon theory}
     \acro  {SCPH+B}[SCPH+B]{self-consistent phonon theory with the bubble diagram}
     \acro  {non-SC}[non-SC]{non self-consistent calculations}
     \acro  {WF}    [WF]    {Wannier function}
     \acro  {MLWF}  [MLWF]  {maximally-localized Wannier function}
     \acrodefplural{MLWF} [MLWFs]  {maximally-localized Wannier functions}
     \acro  {WC}    [WC]    {Wannier center}
     \acrodefplural{WC} [WCs]  {Wannier centers}
     \acro  {BC}    [BC]    {bond center}
     \acrodefplural{BC}    [BCs]  {bond centers}
     \acro  {LDA}   [LDA]   {local density approximation}
     \acro  {PBE}   [PBE]   {Perdew-Bruke-Ernzerhof}
\acro  {CPMD}  [CPMD]  {Car-Parrinello molecular dynamics}
     \acro  {BOMD}  [BOMD]  {Born-Oppenheimer molecular dynamics}
     \acro  {CMD}   [CMD]  {classical molecular dynamics}
     \acro  {BO}    [BO]    {Born-Oppenheimer}
     \acro  {AIMD}  [AIMD]  {\textit{ab initio} molecular dynamics}
     \acro  {MD}    [MD]    {molecular dynamics}
     \acro  {LASSO} [LASSO] {least absolute shrinkage and selection operator}
     \acro  {IFC}   [IFC]   {interatomic force constant}
     \acro  {harm}  [harm]  {harmonic approximation}
     \acro  {TDOS}  [TDOS]  {two phonon density of state}
     \acro  {traj.} [traj.] {trajectory}
     \acro  {lp}    [lp]    {lone-pair}

\end{acronym}

\section{Introduction}\label{sec:intro}

Dielectric properties represent the interaction between electric fields and materials. Optical spectroscopy is a widely used technique to study the structure and dynamics of bulk systems. Simulation of the dielectric properties of materials with strong intermolecular interactions, such as liquid alcohols, is essential for interpreting experimental spectra.

Analyzing how polarization occurs is the key to comprehending the dielectric properties of liquid alcohol. In the liquid phase, the electric field generated by the surrounding molecules distorts the electronic cloud, resulting in different polarization features from the gas phase. Hydrogen bonding is an important intermolecular interaction appearing in polar molecules. Alcohol molecules, consisting of both polar hydroxyl and non-polar alkyl groups, can accept two \acp{H-bond} and only donate one \ac{H-bond}, unlike water molecules that can donate two \ac{H-bond}s. This difference leads to the distinct hydrogen bonding in alcohols compared to water~\cite{shinokita2015Hydrogen,jorge2022Dipole}, thus resulting in the unique dielectric properties of alcohols. The dielectric constant at $\SI{300}{\kelvin}$ reaches $78$ for water, while it is only $33$ for methanol~\cite{davidson1957DIELECTRIC}. In the region below $\SI{1000}{\per\cm}$ (THz region), which is dominated by the intermolecular modes, the dielectric absorption of water has a translational (\ac{H-bond} stretching) peak around $\SI{200}{\per\cm}$ and a libration peak around $\SI{600}{\per\cm}$~\cite{carlson2020Exploring}. Although methanol also has a libration peak around $\SI{700}{\per\cm}$~\cite{guillot1990Investigation,bertie1993Infrared,bertie1994Infrared,bertie1997Infrared}, its lower frequency spectrum is more complex than that of water. THz spectroscopy experiments~\cite{fukasawa2005Relation,sarkar2017Broadband,yomogida2010Dielectric,yomogida2010Comparativec} pointed out three major peaks at around $60$, $120$, and $\SI{270}{\per\cm}$ using the Lorentzian fitting, with the $\SI{120}{\per\cm}$ one being the largest. The $\SI{270}{\per\cm}$ peak is broad and is thought to originate from intermolecular motions~\cite{guillot1990Investigation}. \ac{H-bond} fluctuations, observed in the similar frequency region in experiments~\cite{shirota1997Deuterium,shinokita2015Hydrogen}, are likely associated with these modes. However, the origins of these peaks are not yet fully understood.

After Rahman~\cite{rahman1964Correlations} introduced the \ac{CMD} method in 1964, the dielectric function of a system can be calculated from the dipole auto-correlation function along \ac{CMD} trajectories based on the linear response theory~\cite{kubo1957StatisticalMechanical}. The dipole moment $\vb{M}$ is calculated by assigning an empirical charge $q_i$ to the atom $\vb{r}_i$ as $\vb{M}=\sum_{i}q_{i}\vb{r}_i$. Despite the rather simple description of dipole moments, \ac{CMD} has been successfully used to study dielectric properties. As for THz absorption of methanol, early studies~\cite{matsumoto1990Hydrogen,skaf1993Wave,ladanyi1996Wave,venables2000Structure} identified a significant peak around $\SI{700}{\per\cm}$ to the hydroxyl \ce{H} rotational (librational) motion around the \ce{CO} axis, nearly parallel to the axis of the least inertia. Skaf et al.~\cite{skaf1993Wave,ladanyi1996Wave} quantitatively reproduced lower the $60$ and $\SI{120}{\per\cm}$ absorption peaks, which were imputed to the libration around the second-largest and largest principal inertia axes, respectively~\cite{matsumoto1990Hydrogen,palinkas1991Molecular,venables2000Structure,garberoglio2002Instantaneous,garberoglio2001Instantaneous}. Additionally, considering the induced dipole moments improved the agreement with experiments~\cite{marti1995Hydrogen}. Torii~\cite{torii2023Intermolecular} demonstrated that the induced dipole moments significantly contributed to the low-frequency peaks, while the rotation of the permanent molecular dipoles accounted for the $\SI{700}{\per\cm}$ peak of methanol by incorporating the induced dipole moments through \ac{H-bond} stretching.

Although empirical force field approaches have been widely employed in studies on liquid alcohols, qualitative agreement with experimental values of the dielectric constant~\cite{shilov2015Molecular,kulschewski2013Molecular} and the dielectric function~\cite{skaf1993Wave,venables2000Structure,torii2023Intermolecular} remains challenging due to the lack of explicit many-body induced polarization effects.

Car and Parrinello developed the \ac{AIMD} method~\cite{car1985Unified}, in which the atomic forces are calculated directly from first principle electronic structure calculations without empirical parameters. Furthermore, the modern theory of polarization~\cite{resta1992Theorya,king-smith1993Theory} can express the dipole moment in a quantum mechanics manner through the \ac{MLWF} method~\cite{marzari1997Maximally,marzari2012Maximally}, where the electronic contribution to the dipole moment is evaluated from the $-2e$ point charges at the centers of \acp{MLWF}, known as \acp{WC}. Accurate estimation of dynamics and polarization has led to numerous studies on liquid alcohols focusing on structure~\cite{tsuchida1999Densityfunctional,tsuchida2008Initio,handgraaf2003Initio,mcgrath2011Liquid,mcgrath2011Vapor}, hydrogen bonding~\cite{pagliai2003Hydrogen,yadav2012First,jindal2020Geometry}, dipole moments~\cite{hait2018How,jorge2022Dipole}, and dielectric properties~\cite{handgraaf2004Densityfunctional,woods2005Influence,sieffert2013Liquid,wang2017Initio}. These studies have been elucidating the influence of the local atomic environment on dielectric properties. The average molecular dipole moment in the liquid methanol calculated from \ac{AIMD} ranged from $\SI{2.5}{\mathrm{D}}$ to $\SI{2.9}{\mathrm{D}}$~\cite{pagliai2003Hydrogen,handgraaf2003Initio,handgraaf2004Densityfunctional,sieffert2013Liquid}, which was larger than those obtained from force-field calculations~\cite{haughney1987Moleculardynamics}. Pagliai et al.~\cite{pagliai2003Hydrogen} showed that the number of \ac{H-bond}s strongly influenced the molecular dipole moments. Wang et al.~\cite{wang2017Initio} successfully reproduced the experimental dielectric function of liquid methanol using \ac{BOMD}, showing significant improvement compared to the \ac{CPMD} calculation, which suffered from a large redshift in the spectrum around $\SI{3000}{\per\cm}$~\cite{handgraaf2004Densityfunctional}. For THz dielectric spectra, studies on liquid water unveiled the importance of the polarization of \ac{WC}s due to intermolecular interactions~\cite{sharma2005Intermolecularb,chen2008Role,carlson2020Exploring}. In contrast, the THz spectra calculation of liquid methanol has not been done so often~\cite{woods2005Influence}.

Recently, \ac{ML} methods have been employed to express \textit{ab initio} potential energy surfaces as a function of nuclear coordinates~\cite{behler2007Generalized,behler2015Constructing,wang2018DeePMDkit,zeng2023DeePMDkit,artrith2016Implementation}. These methods preserve the accuracy of AIMD while improving its efficiency. To calculate dielectric properties using this technique, however, we also need \ac{ML} models to predict the dipole moment from atomic positions. To this end, Gastegger et al.~\cite{gastegger2017Machine} constructed a model that learns the environment-dependent atomic partial charges, which vary depending on the positions of neighborhood atoms, and thus effectively describes the polarization effect of electrons in gas phase molecules. This scheme has been successfully applied to clusters~\cite{beckmann2022Infrared,bereau2018Noncovalenta} and liquid structures~\cite{metcalf2021ElectronPassing,bleiziffer2018Machinea} and has been utilized to incorporate long-range Coulomb interactions into \ac{ML} molecular dynamics~\cite{yao2018TensorMol0,unke2019PhysNet}. This atomic charge model has been adapted with \ac{ML} models to learn dipole or multipole moments using ordinary \ac{ML} methods~\cite{kapil2020Inexpensive,veit2020Predicting,hou2020Dielectric} or graph-neural networks~\cite{glick2021Cartesian,thurlemann2022Learning,zhong2023General}.

An alternative approach is to directly handling \acp{WC}. Several authors constructed ML models to predict the average position of the \acp{WC} in a water molecule, called the Wannier centroid, and successfully reproduced dielectric properties~\cite{krishnamoorthy2021Dielectric,hou2020Dielectric,zhang2020Deep,sommers2020Raman,zhang2022Deep}. To accurately predict vector quantities such as \acp{WC}, the models must be rotationally equivariant when the reference coordinate system is rotated. Zhang et al.~\cite{zhang2020Deep,sommers2020Raman} addressed this requirement by using two networks, referred to as embedding and fitting networks. While a water molecule only has four \acp{WC}, larger molecules contain a larger number of \acp{WC}, making predicting the centroids more difficult. To overcome this challenge, we aim to extend this method to construct \ac{ML} models for each \ac{WC} rather than the Wannier centroid.

In this work, we propose a versatile \ac{ML} model to predict \ac{WC}s applicable to molecular liquids. We assign \ac{WC}s to chemical bonds between atoms and use deep neural networks to predict the position of the \ac{WC} for each bond. WCs are collected by DFT calculations and used as training data for \ac{ML}. To ensure the equivariance of the models, we followed the method by Zhang et al.~\cite{zhang2020Deep}. We applied our method to calculate the dielectric properties of liquid methanol and ethanol. The \ac{ML} models accurately predict the molecular dipole moments and reproduce the experimental dielectric properties well in the THz region, combined with \ac{AIMD} trajectories.

\section{Theory}\label{sec:theory}

\subsection{Dipole Moments}
\begin{figure}[tb]
\centering

 \begin{subcaptionblock}{0.5\linewidth}
  \subcaption{Methanol} \end{subcaptionblock}\hfill
 \begin{subcaptionblock}{0.5\linewidth}
  \subcaption{Ethanol} \end{subcaptionblock}\hfill
  \includegraphics[width=\linewidth]{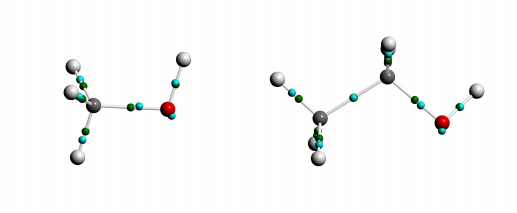}

\caption{Illustration of methanol (left) and ethanol (right) molecules. Oxygen, carbon, and hydrogen are represented by red, gray, and white spheres, respectively. Bond centers (BCs) and \ac{WC}s are represented by small green and blue spheres. Only one blue circle is visible for the O \ac{lp}, as it is the center of gravity of the two \acf{WC}.}
\label{Fig:wcs}
\end{figure} \begin{figure}[tb]
 \centering
\centering
\begin{subcaptionblock}{0.5\linewidth}
   \subcaption{Single bond} 
   \adjustbox{right}{\includegraphics[width=\linewidth]{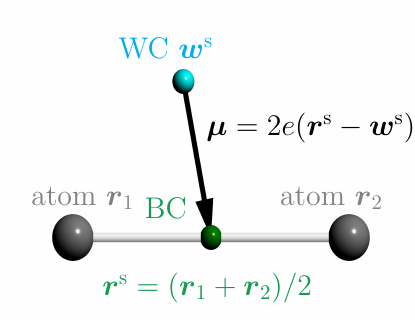}}
\end{subcaptionblock}\hfill
\begin{subcaptionblock}{0.5\linewidth}
   \subcaption{Lone pair}
   \adjustbox{right}{\includegraphics[width=\linewidth]{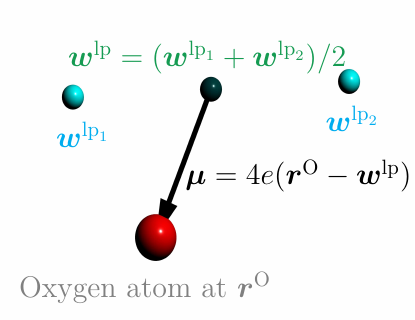}}
\end{subcaptionblock}
  \caption{Schematic image of dipole moments for a single bond and lone pair. $\vb{r}$ represents the position of BCs.}
\label{Fig:dipole_decompose}
\end{figure} \begin{figure*}[tb]
 \centering
  \includegraphics[width=\linewidth]{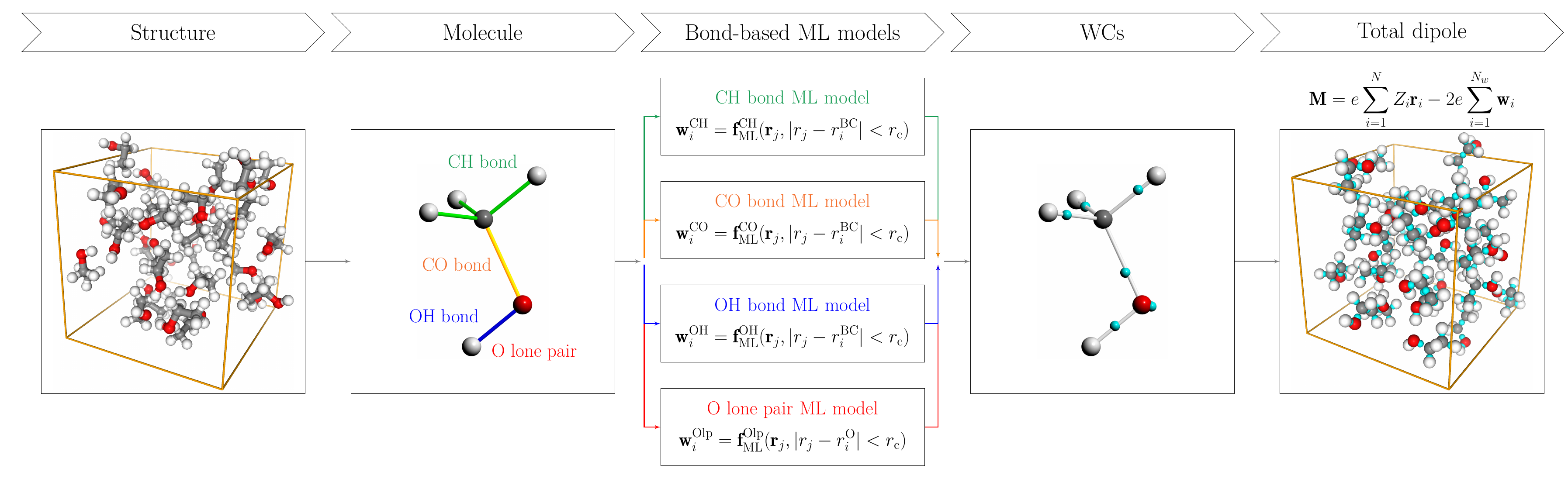}
\caption{Workflow to calculate the total dipole moment from bond-based \ac{ML} dipole models. We create \ac{ML} models for each bond species to predict the \ac{WC} assigned to the bond. Predicting all \ac{WC}s in the system, we obtain the total dipole moment according to \cref{EqWF}.}
\label{Fig:mlworkflow}
\end{figure*}

Evaluating dielectric properties requires the total dipole moments of the system during \ac{MD} simulations. In the \ac{MLWF} method~\cite{marzari1997Maximally,marzari2012Maximally}, the total dipole moment of a bulk periodic system consists of the ion-core and the valence-electronic parts. The ion-core part can be calculated from ion-core charges and positions as usual, whereas the valence-electronic contribution is evaluated from the $-2e$ point charges at $i$-th \ac{WC} $\vb{w}_{i}$ as
\begin{align}\label{EqWF}
 \vb{M} = \vb*{M}_{\mathrm{atom}} + \vb*{M}_{\mathrm{el}} = e\sum_{i=1}^{N} Z_{i}\vb{r}_{i} -2e\sum_{i=1}^{N_w}\vb{w}_{i},
\end{align}
where $Z_{i}$ is the ion-core charge of atoms, and $\vb{r}_i$ denotes the nuclear position. $N$ and $N_w$ are the number of atoms and \acp{WC}, respectively. The coefficient of $-2$ comes from the degeneracy of spin degrees of freedom. We use a valence-only pseudopotential approach so that the \acp{WC} correspond to the valence electrons, and ion-core charges $eZ_{i}$ are the sums of the charges of the bare nuclei and the frozen core electrons.

Our central idea is to rewrite \cref{EqWF} into a chemical bond-based expression. We assume that \acp{WC} are close enough to the centers of chemical bonds, called \acp{BC}, or ion-cores in case of electron lone pairs, so that each \ac{WC} can be clearly allocated to the corresponding \ac{BC} or ion-core one by one, which is true in ordinary molecular systems. The examples of methanol and ethanol, shown in \cref{Fig:wcs}, illustrates that the \ac{WC}s are well localized on the chemical bonds or the O lone pairs.

To clarify the discussion, we consider the system composed of oxygens, carbons, and hydrogens with single bonds and oxygen lone pairs in this paper. The extension to systems with double bonds or triple bonds is straightforward. We can associate each \ac{WC} with a specific chemical bond or a lone pair, i.e., a single bond has one \ac{WC} at $\vb{w}^{\mathrm{s}}$, and an oxygen atom has two lone pair \ac{WC}s at $\vb{w}^{\mathrm{lp}_{1}}$ and $\vb{w}^{\mathrm{lp}_{2}}$. Therefore, the electronic term in \cref{EqWF} can be recast to the bond-based form as
\begin{align}
 \vb*{M}_{\mathrm{el}} &= -2e\sum_{i=1}^{N_{\mathrm{single}}} \vb{w}^{\mathrm{s}}_{i}-2e\sum_{i=1}^{N_{\mathrm{lp}}}\left(\vb{w}^{\mathrm{lp}_{1}}_{i}+\vb{w}^{\mathrm{lp}_{2}}_{i}\right)\label{211301_7Feb24} \\
 &= -2e\sum_{i=1}^{N_{\mathrm{single}}} \vb{w}^{\mathrm{s}}_{i}-4e\sum_{i=1}^{N_{\mathrm{lp}}}\vb{w}^{\mathrm{lp}}_{i},\label{150814_8Feb24}
\end{align}
where $N_{\mathrm{single}}$ and $N_{\mathrm{lp}}$ are the number of single bonds and lone pairs, respectively. $N_{\mathrm{lp}}$ is also identical to the number of oxygen atoms in our case. $\vb{w}^{\mathrm{lp}}_{i}=\left(\vb{w}^{\mathrm{lp}_{1}}_{i}+\vb{w}^{\mathrm{lp}_{2}}_{i}\right)/2$ is the center of mass of two lp \acp{WC}. Henceforth, we also treat $\vb{w}^{\mathrm{lp}}_{i}$ as \acp{WC}. \Cref{150814_8Feb24} is the most important equation, representing the one-to-one correspondence between each chemical bond (or lp) and \acp{WC}.

To further simplify the equation, we devise a similar decomposition for the ion-core term in \cref{EqWF}. A carbon atom has four bonds, a hydrogen atom has one bond, and an oxygen atom has one bond and two lone pairs. Considering that atoms supply the associated single bonds with $+e$ charge and lone pairs with $+2e$ charges, we can also express the atomic contribution in \cref{EqWF} in the bond-based form as
\begin{align}
 \vb{M}_{\mathrm{ion}} &= e\sum_{i=1}^{N_\mathrm{single}}\left(\vb{r}^{1}_{i}+\vb{r}^{2}_{i}\right) + 4e\sum_{i=1}^{N_\mathrm{lp}}\vb{r}^{\ce{O}}_{i}\label{211311_7Feb24} \\
        &= 2e\sum_{i=1}^{N_\mathrm{single}}\vb{r}^{s}_{i} + 4e\sum_{i=1}^{N_\mathrm{lp}}\vb{r}^{\ce{O}}_{i},
\end{align}
where $\vb{r}^{\lambda}_{i}\, (\lambda=1,2)$ denotes the atomic position associated with the bond $i$, and $\vb{r}^{s}_{i}=\left(\vb{r}^{1}_{i}+\vb{r}^{2}_{i}\right)/2$ are the \ac{BC}s. $\vb{r}^{O}_{i}$ is the position of the oxygen $i$. Adding \cref{211301_7Feb24,211311_7Feb24}, we finally reach the bond-based expression of the total dipole moment as 
\begin{align}
 \vb{M} &= 2e\sum_{i=1}^{N_\mathrm{single}}\left(\vb{r}^{s}_{i}-\vb{w}^{\mathrm{s}}_{i}\right)+4e\sum_{i=1}^{N_\mathrm{lp}}\left(\vb{r}^{\ce{O}}_{i}-\vb{w}^{\mathrm{lp}}_{i}\right) \\
 &=\sum_{i=1}^{N_\mathrm{single}}\vb*{\mu}_{i}^{\mathrm{single}}+\sum_{i=1}^{N_\mathrm{lp}}\vb*{\mu}_{i}^{\mathrm{lp}}.\label{Eq:bd}
\end{align}
We call $\vb*{\mu}_{i}^{\mathrm{single}}=2e\left(\vb{r}^{s}_{i}-\vb{w}^{\mathrm{s}}_{i}\right)$ and $\vb*{\mu}_{i}^{\mathrm{lp}}=4e\left(\vb{r}^{\ce{O}}_{i}-\vb{w}^{\mathrm{lp}}_{i}\right)$ bond dipoles, which are just the relative vectors from \ac{BC}s (or oxygens) to \acp{WC}, as shown in \cref{Fig:dipole_decompose}. In the following, we equate bond dipoles with WCs. \Cref{Eq:bd} shows that the total dipole moment is decomposed into bond and lone pair components. Therefore, with \ac{ML} models that predict the bond dipole for each bond and lone pair from neighboring atomic coordinates $\{\vb{R}\}$ as $\vb*{\mu}=\vb*{f}(\{\vb{R}\})$, we can infer the total dipole moments, as shown in \cref{Fig:mlworkflow}.

\subsection{Model Construction}

We construct a neural network vector function $\vb{f}$ that predicts the bond dipole $\vb*{\mu}_i$, given the coordinates of the neighborhood atoms $\vb{r}_k\in N_{i}$ around the \ac{BC} $\vb{r}^{\mathrm{s}}_{i}$ or lone pair atom $\vb{r}^{\mathrm{\ce{O}}}_{i}$ within the cutoff radius $r_c$ as $\vb*{\mu}_i = \vb{f}(\left\{\vb{r}_k\in N_{i}\right\})$.

We followed the method proposed by Zhang et al.~\cite{zhang2020Deep} so that $\vb{f}$ satisfies the translational, permutational, and rotational symmetry. To preserve translational symmetry, we make a relative coordinate from a target \ac{BC} $\vb{r}^{\mathrm{s}}_{i}$ or lone pair atom $\vb{r}^{\ce{O}}_{i}$ as $\vb{r}'_{ki}=\vb{r}_{k}-\vb{r}^{\mathrm{s}}_{i}=(x_{ki}, y_{ki}, z_{ki})$ or $\vb{r}'_{ki}=\vb{r}_k-\vb{r}^{\ce{O}}_{i}=(x_{ki}, y_{ki}, z_{ki})$, respectively. Then, we introduce the following cutoff function $s(r)$ as \begin{align}
  s(r)=
  \begin{cases}
   \frac{1}{r} & r<r_{\mathrm{c}_{0}} \\
   \frac{1}{r}\left\{\frac{1}{2}\cos\left(\pi \frac{r-r_{\mathrm{c}_{0}}}{r_{\mathrm{c}_{0}}-r_{\mathrm{c}}}\right)+\frac{1}{2}\right\}& r_{\mathrm{c}_{0}}<r<r_{\mathrm{c}} \\
   0& r_{\mathrm{c}} < r.
  \end{cases}
 \end{align}
We describe the atomic coordinates $\vb{r}'_{ik}$ with a four component vector 
\begin{align}
\vb{q}_{ik}&= (q^{1}_{ik},q^{2}_{ik},q^{3}_{ik},q^{4}_{ik}) \\
           &= (s(r'_{ik}),s(r'_{ik})x'_{ik},s(r'_{ik})y'_{ik},s(r'_{ik})z'_{ik}).
\end{align}
The $N_i$ by $4$ matrix $\vb{Q}=(Q_{k\lambda})=(q^{\lambda}_{ik})$ represent the set of coordinates $\{\vb{q}_{ik}\}$ in a neighborhood, where $\lambda$ is the Cartesian index. In the actual calculation, $\vb{q}_{ik}$ are ordered by atomic species in ascending order with respect to $r'_{ki}$.

Next, we introduce two deep neural networks (DNN), an embedding DNN and a fitting DNN to ensure permutational invariance and rotational covariance. The embedding DNN $\vb{E}(\{s(r')\})$ is the mapping from the set $\{q^{1}_{ik} | k=1,\cdots,N_{i}\}$ onto the matrix with $M$ rows and $N_i$ columns. To reduce the computational cost, we define the truncated embedding matrix $\vb{E}'$ formed by the embedding matrix's first $M'(<M)$ columns. Using $\vb{E}$ and $\vb{E}'$, we construct the feature matrix $\vb{D}$ of dimension $M\times M'$ as
\begin{align}
 \vb{D}_{i} = (\vb{E}\vb{Q})(\vb{E}'\vb{Q})^{T} = \vb{E}\vb{Q}\vb{Q}^{T}\vb{E}'^{T}.
\end{align}
In the feature matrix, translational and rotational symmetries are preserved by the matrix product of $\vb{Q}\vb{Q}^{T}$.

The fitting DNN $\vb{F}(\vb{D})$ maps $\vb{D}_{i}$ onto $M$ outputs $F_{j} (j=1,2,\cdots, M)$, which are finally used to calculate $\vb*{\mu}_{i}=(\mu^{1}_{i},\mu^{2}_{i},\mu^{3}_{i})$ with the last three columns of $\vb{T}=\vb{E}\vb{Q}$,
\begin{align}
 \mu_{i}^{\lambda} = \sum_{j=1}^{M} F_j(\vb{D}) T_{j,\lambda+1} \, (\lambda=1,2,3).
\end{align}

We define the loss function as
\begin{align} L = \frac{1}{n_{\mathrm{b}}}\sum_{i=1}^{n_{\mathrm{b}}}\left|\vb*{\mu}^{\mathrm{p}}_{i}-\vb*{\mu}^{\mathrm{t}}_{i}\right|^2,\label{231308_29Feb24}
\end{align}
where $\vb*{\mu}^{\mathrm{p}}_{i}$ and $\vb*{\mu}^{\mathrm{t}}_{i}$ are the predicted and DFT bond dipole moments of the $i$-th data, respectively, and $n_{\mathrm{b}}$ is the batch size. We train different \ac{ML} models for each chemical bond species according to \cref{231308_29Feb24}.

\subsection{The dielectric properties}
\begin{figure}[tb]
\centering
  \includegraphics[width=0.6\linewidth]{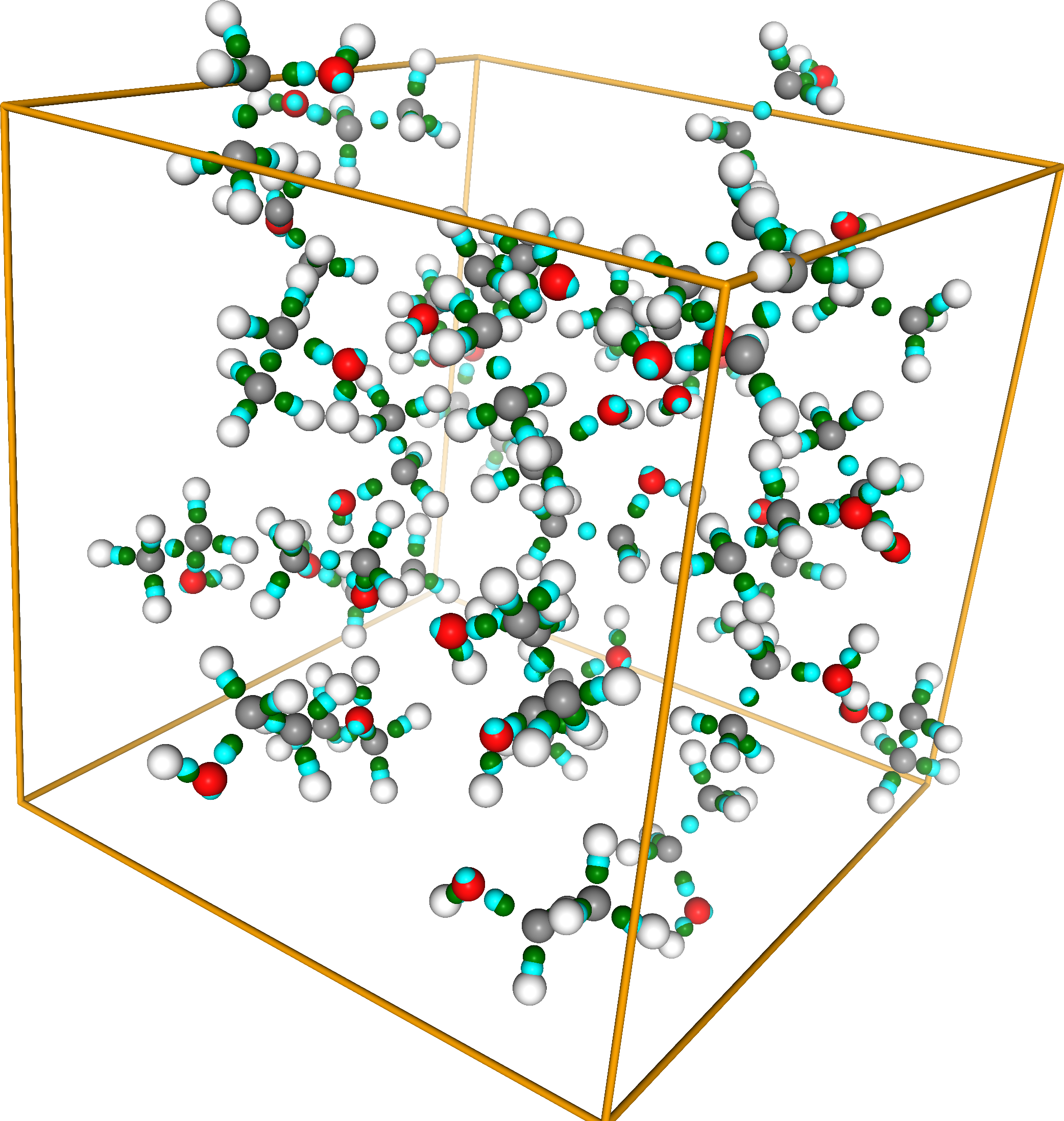}
  \caption{Snapshot of MD simulation with \ac{ML} dipole moments for liquid ethanol containing $32$ molecules. The coordinates of the \ac{BC}s (dark green) are computed at each MD step, and the \ac{ML} models predict the WCs (light blue). Hydrogen, oxygen, and carbon atoms are shown in white, red, and gray, respectively.}
\label{Fig:md_snapshot}
\end{figure}  

\begin{figure}[tb]
 \centering
 \includegraphics[width=\linewidth]{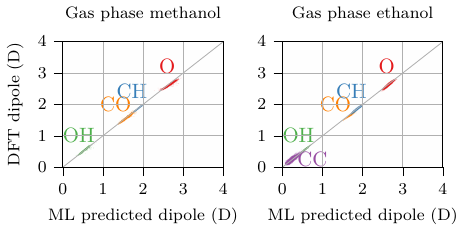}
 \caption{Learning accuracy of \ac{ML} dipole models for gas phase (left) methanol and (right) ethanol, where the $x$-axis represents the value predicted by ML and the $y$-axis represents the value calculated by DFT. There are four models for methanol and five for ethanol. The data are taken from both training and test data. The blue, orange, green, red, and purple dots represent the absolute values of the CH, CO, OH, O, and CC bond dipole moments, respectively.}
\label{fig:lr_gas_met}
\end{figure} 

\begin{table}[tb] 
\centering
\caption{ RMSE $[\mathrm{D}]$ of \ac{ML} dipole models for methanol and ethanol, calculated from validation data. The \ce{CC} bond model exists only for ethanol.}
\input{lr.tex} \label{table:lr}
\end{table} 

The dielectric function at angular frequency $\omega$ of the external electric field is a complex quantity written as
\begin{align}
\varepsilon(\omega)=\varepsilon'(\omega)-i\varepsilon''(\omega),
\end{align}
where $\varepsilon'$ and $\varepsilon''$ are the real and imaginary parts, respectively. We consider the relative dielectric function, and all $\varepsilon$ are dimensionless. From the linear response theory, the dielectric function $\varepsilon(\omega)$ of an isotropic and homogeneous fluid in MD simulations with periodic boundary conditions is given by~\cite{neumann1983Calculation,neumann1984Consistent,neumann1984Computer,sharma2007Dipolara}
\begin{align} \varepsilon(\omega)=\varepsilon^{\infty} -\frac{1}{3k_{\mathrm{B}}T} \int_{0}^{\infty}\left(\frac{\dd \expval{\tilde{\vb{M}}(0)\cdot \tilde{\vb{M}}(t)}}{\dd t}\right)e^{-i\omega t}\dd t,\label{013609_20Feb24}
\end{align}
where $\expval{}$ denote the canonical ensemble averages, and $\tilde{\vb{M}}(t)= \vb{M}(t)-\expval{\vb{M}}$ is the zero-mean dipole moment along a \ac{MD} trajectory. $\varepsilon^{\infty}$ is the high-frequency dielectric constant, $k_{\mathrm{B}}$ is the Boltzmann constant, $T$ is the temperature, and $V$ is the volume of the simulation cell. Using the static dielectric constant
\begin{align} \varepsilon^{0} = \varepsilon(0) = \varepsilon^{\infty}+\frac{1}{3k_{\mathrm{B}}TV}\left(\expval{\vb{M}^2}-\expval{\vb{M}}^2\right),\label{171945_26Feb24}
\end{align}
we rewrite \cref{013609_20Feb24} as
\begin{align} \frac{\varepsilon(\omega)-\varepsilon^{\infty}}{\varepsilon^{0}-\varepsilon^{\infty}} = \int_{0}^{\infty}\left(-\frac{\dd \Phi(t)}{\dd t}\right)e^{-i\omega t}\dd t,\label{141231_5Feb24}
\end{align}
where $\Phi(t)$ is the Fourier transform of the normalized autocorrelation function of the dipole moment $\vb*{\tilde{M}}(t)$,
\begin{align}
 \Phi(t)=\frac{\expval{\tilde{\vb{M}}(0)\cdot \tilde{\vb{M}}(t)}}{\expval{\tilde{\vb{M}}^2}}.
\end{align}
To avoid evaluating the derivative of the auto-correlation function, we adopt the alternative form derived from the integration by parts of \cref{141231_5Feb24}~\cite{cardona2018Molecular}
\begin{align}
 \frac{\varepsilon(\omega)-\varepsilon^{\infty}}{\varepsilon^{0}-\varepsilon^{\infty}} = 1-i\omega \int_{0}^{\infty} \Phi(t)e^{-i\omega t}\dd t .\label{183933_8Feb24}
\end{align}

The complex refractive index $\hat{n}(\omega)$ is another important optical quantity defined as 
\begin{align}
\hat{n}(\omega) = n(\omega)-i\kappa(\omega),
\end{align}
where the real part $n(\omega)$ is the refractive index, and the imaginary part $\kappa(\omega)$ is the optical extinction coefficient. The complex dielectric constant and the complex refractive index are related as
\begin{align}
 \varepsilon(\omega) = \hat{n}(\omega)^2.
\end{align}
Therefore, the imaginary part of the dielectric function is 
\begin{align}
 \varepsilon''(\omega) = 2n(\omega)\kappa(\omega).
\end{align}
Absorption coefficient per unit length $\alpha(\omega)$ is defined through Lambert-Beer's law~\cite{hecht2017Optics}: \begin{align}
\alpha(\omega) = \frac{2\omega}{c}\kappa(\omega),
\end{align}
where $c$ is the speed of light in vacuum. Absorption spectrum $\alpha(\omega)n(\omega)$ is associated with $\varepsilon''(\omega)$ via the equation
\begin{align} \alpha(\omega)n(\omega)=\frac{2\omega k(\omega)}{c}n(\omega)=\frac{\omega}{c}\varepsilon''(\omega).\label{195505_29Feb24}
\end{align}
To analyze the spectra, we write the total dipole moment as the aggregate of the $N$ individual molecular dipole moments, $\vb{M}(t) = \sum_{i=1}^{N}\vb*{\mu}^{\mathrm{mol}}_{i}(t)$, where $\vb*{\mu}^{\mathrm{mol}}_{i}(t)$ is the dipole moment of the $i$-th molecule. Then, the autocorrelation function appearing in \cref{013609_20Feb24} can be decomposed as
\begin{align}\expval{\vb{M}(0)\cdot \vb{M}(t)} &= \sum_{i=1}^{N}\expval{\vb*{\mu}^{\mathrm{mol}}_{i}(0)\cdot \vb*{\mu}^{\mathrm{mol}}_{i}(t)} \notag\\
 &+\sum_{i\neq j}\expval{\vb*{\mu}^{\mathrm{mol}}_{i}(0)\cdot \vb*{\mu}^{\mathrm{mol}}_{j}(t)}.\label{195111_29Feb24}
\end{align}
The first term is the autocorrelations of the dipole moments of single molecules, and the second is cross-correlations among the dipole moments of different molecules~\cite{chen2008Role}. Substituting \cref{195111_29Feb24} into \cref{013609_20Feb24}, we acquire the self and collective component of the absorption as
\begin{align}
 \varepsilon''_{\mathrm{self}}(\omega) &= -\frac{1}{3k_{\mathrm{B}}T}\int_{0}^{\infty}\dv{t}\left(\sum_{i=1}^{N}\expval{\vb*{\mu}^{\mathrm{mol}}_{i}(0)\cdot \vb*{\mu}^{\mathrm{mol}}_{i}(t)}\right)e^{-i\omega t}\dd t, \\
 \varepsilon''_{\mathrm{coll}}(\omega) &= -\frac{1}{3k_{\mathrm{B}}T}\int_{0}^{\infty}\dv{t}\left(\sum_{i\neq j}\expval{\vb*{\mu}^{\mathrm{mol}}_{i}(0)\cdot \vb*{\mu}^{\mathrm{mol}}_{j}(t)}\right)e^{-i\omega t}\dd t,
\end{align}
yielding 
\begin{align} \alpha(\omega)n(\omega)= \frac{\omega}{c}\varepsilon''_{\mathrm{self}}(\omega)+\frac{\omega}{c}\varepsilon''_{\mathrm{coll}}(\omega).\label{180524_26Feb24}
\end{align}

\subsection{The velocity auto-correlation function}
\begin{figure*}[htb]\centering
\captionsetup[subfigure]{font={bf,large}, skip=1pt, margin=-0.7cm,justification=raggedright, singlelinecheck=false}
 \begin{subcaptionblock}{0.45\linewidth}
  \subcaption{Bond dipole moment} \centering
 \adjustbox{right}{\includegraphics[width=0.7\textwidth]{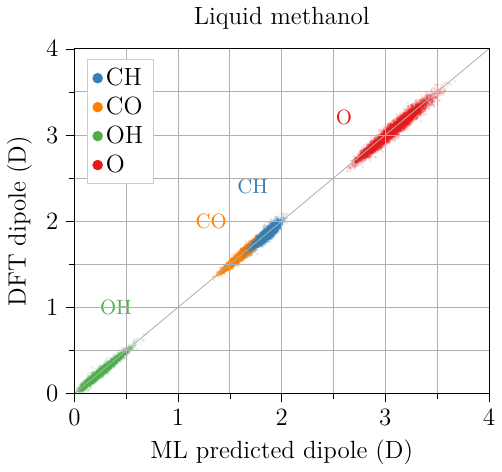}}
  \end{subcaptionblock}
\begin{subcaptionblock}{0.45\linewidth}
  \captionsetup[subfigure]{labelformat=empty}
   \centering
   \adjustbox{left}{\includegraphics[width=0.7\textwidth]{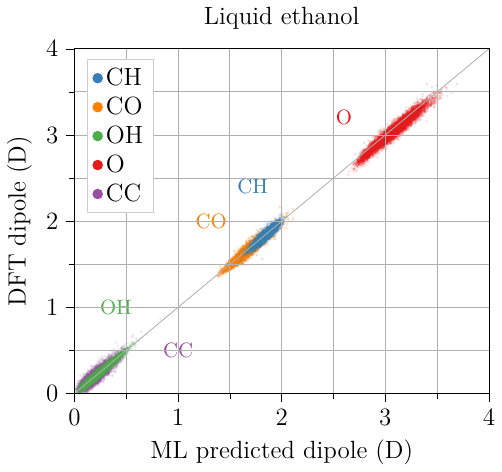}}
\end{subcaptionblock}

\begin{subcaptionblock}{0.45\linewidth}
 \captionsetup[subfigure]{font={bf,large}, skip=1pt, margin=-0.7cm,justification=raggedright, singlelinecheck=false}
  \subcaption{Molecular dipole moment} \centering
  \adjustbox{right}{\includegraphics[width=0.7\textwidth]{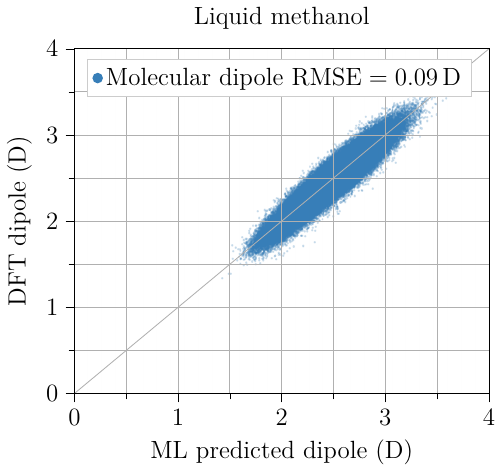}}
  \end{subcaptionblock}
\begin{subcaptionblock}{0.45\linewidth}
  \captionsetup[subfigure]{labelformat=empty}
  \centering
  \adjustbox{left}{\includegraphics[width=0.7\textwidth]{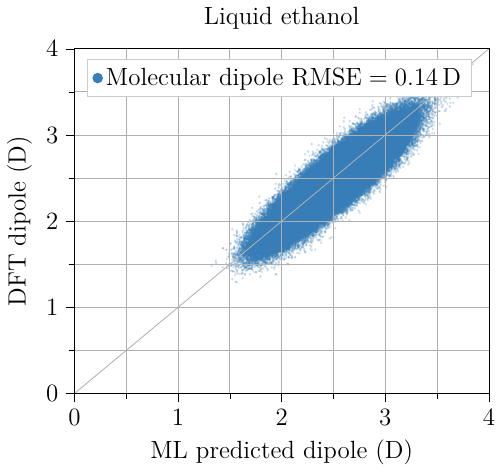}}
 \end{subcaptionblock} 
\caption{(a) Learning accuracy of the bond dipole moments of (left) methanol and (right) ethanol, where the $x$-axis represents the value predicted by ML and the $y$-axis represents the value calculated by DFT. $10000$ validation data are plotted.(b) Learning accuracy of the molecular dipole moments calculated from from the bond dipole model for liquid methanol and ethanol.}
\label{fig:lr_met}
 \end{figure*} The \ac{VACF} of atom $i$ is defined as
\begin{align}
 v^{\mathrm{ACF}}_i(t) = \expval{\vb*{v}_{i}(0)\cdot\vb*{v}_{i}(t)},
\end{align}
where $\vb*{v}_{i}(t)$ is the atomic velocity at time $t$. The total \ac{VACF} is the average of atomic \ac{VACF} as
\begin{align}
 V^{\mathrm{ACF}}(t) = \frac{1}{N}\sum_{i=1}^{N}v^{\mathrm{ACF}}_i(t),
\end{align}
where $N$ is the number of atoms in the system. We can calculate the \ac{VACF} for each atomic species by restricting the summation. The vibrational density of states is the Fourier transformation of \ac{VACF} as
\begin{align}
 D(\omega) = \int_{-\infty}^{\infty} V^{\mathrm{ACF}}(t)e^{-i\omega t}\dd t,\label{063801_26Feb24}
\end{align}
which gives the spectral information of atomic movements.

The angular velocity auto-correlation function of atom $i$ belonging to molecule $a$, defined as
\begin{align}
 \omega^{\alpha,\mathrm{ACF}}_i(t) = \expval{\omega^{\alpha}_{i}(0)\omega^{\alpha}_{i}(t)},\label{063814_26Feb24} 
\end{align}
represents the intramolecular rotational motions, where $\alpha$ is the Cartesian index. $\vb*{\omega}_{i}(t)$ is the atomic angular velocity at time $t$, which can be calculated from the relative angular momentum $\vb{L}_{i}(t)$ of atom $i$ from the center of mass of the molecule ${a}$ and the inertia tensor $I_{a}$ of the molecule $a$ as
\begin{align}
 \vb*{L}_i(t) = I_{a}(t)\vb*{\omega}_{i}(t).
\end{align}
The eigenvectors $\vb*{e}^{\alpha}_{a}(t)$ and eigenvalues $I^{\alpha}_{a}(t)$ of the inertia tensor are called the principal axes of inertia and principal moments of inertia, respectively. The three principal axes of inertia are frequently used as instantaneous molecular coordinates~\cite{carlson2020Exploring,torii2023Intermolecular}. In this frame, the angular velocity is simply the angular momentum divided by the principal moment of inertia. By analogy with \cref{063801_26Feb24}, the Fourier transform of \cref{063814_26Feb24} gives the spectrum of the rotational motion of the molecules:
\begin{align}D^{\alpha}(\omega) = \int_{-\infty}^{\infty} \frac{1}{N}\sum_{i=1}^{N}\omega^{\alpha,\mathrm{ACF}}_{i}(t)e^{-i\omega t}\dd t.\label{021406_28Feb24}
\end{align}

\section{Results and Discussion}

\subsection{Computational details}

 \begin{figure}[tb]\centering
\captionsetup[subfigure]{font={bf,large}} \begin{subcaptionblock}{0.48\linewidth}
\adjustbox{right}{\includegraphics[width=\textwidth]{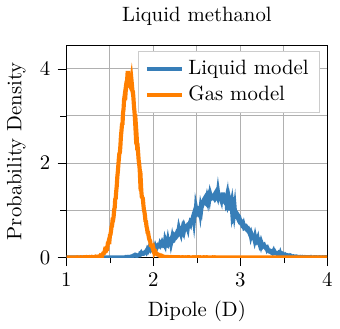}}
\end{subcaptionblock}\hfill
   \begin{subcaptionblock}{0.48\linewidth}
\adjustbox{right}{\includegraphics[width=\textwidth]{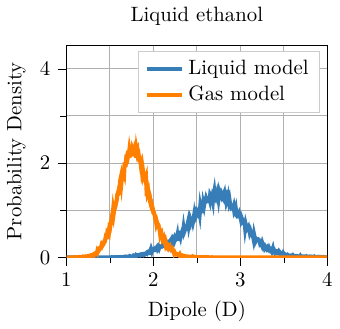}}
  \end{subcaptionblock}
 \caption{Molecular dipole moments of (left) methanol and (right) ethanol in the liquid phase predicted from gas models (orange) and liquid models (blue). Hundred thousand molecules are randomly sampled from $\SI{10}{\pico\second}$ \ac{BOMD} calculations.}
\label{fig:dipole_gas_liquid}
\end{figure}  

\begin{table*}[bt]
\centering
\caption{Molecular dipole moments $\mu^{\mathrm{mol}}$ and dielectric constants $\varepsilon^{0}$ for methanol and ethanol obtained from \ac{CMD} and \ac{AIMD} calculations at $\SI{298}{\kelvin}$ accompanied with experimental data of the gas phase dipole moment, dielectric constant, refractive index $n$, and density $\rho$. $\mu^{\mathrm{mol}}_{\mathrm{G}}$ and $\mu^{\mathrm{mol}}_{\mathrm{L}}$ are the average dipole moments calculated from gas and liquid models, respectively, from the hundred thousand molecular structures sampled from a $\SI{10}{\pico\second}$ \ac{BOMD} trajectory. $\Delta\mu^{\mathrm{mol}} = \mu^{\mathrm{mol}}_{\mathrm{L}}-\mu^{\mathrm{mol}}_{\mathrm{G}}$ is the difference between liquid and gas calculations. Dipoles are in Debye and densities are in $\si{\g/\cm^3}$.}
\input{dipole.tex}
\label{table:dipole}
\end{table*}  

\begin{figure}[tb]\centering
\includegraphics[width=\linewidth]{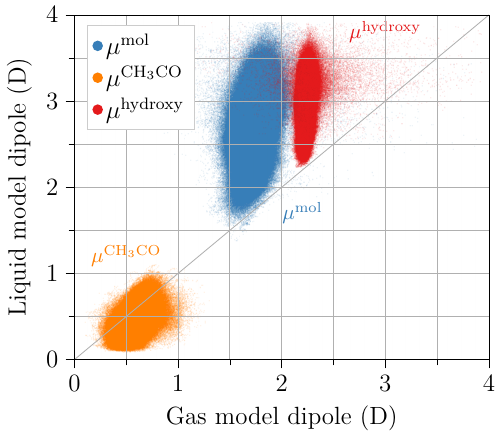}
\caption{Calculated bond dipole moments of $\vb*{\mu}^{\mathrm{\ce{CH3CO}}}=\vb*{\mu}^{\mathrm{\ce{CH3}}}+\vb*{\mu}^{\mathrm{\ce{CO}}}$ (orange), $\vb*{\mu}^{\mathrm{hydroxy}}=\vb*{\mu}^{\mathrm{\ce{OH}}}+\vb*{\mu}^{\mathrm{\ce{O}lp}}$ (red), and $\vb*{\mu}^{\mathrm{mol}}$ (blue). The $x$ and $y$ axes represent the predictions by the gas and liquid models, respectively. Hundred thousand molecules are randomly sampled from $\SI{10}{\pico\second}$ \ac{BOMD} calculations.}
\label{fig:dipole_gas_liquid_2}
 \end{figure} 

 \begin{figure}[tb]\centering
  \includegraphics[width=\linewidth]{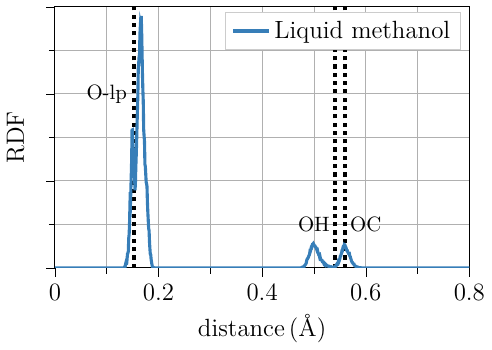}
 \caption{The radial distribution function from O atoms to WCs in liquid methanol. In the case of a lone pair, it is the center of gravity of the two WCs. The three peaks correspond to the O lone pair, the second to the WCs of the OH bond, and the third to the WCs of the OC bond, in increasing order of distance. The vertical lines indicate the peak positions in the gas phase.}
\label{fig:o_wc_rdf}
 \end{figure} 

We used the CPMD package~\cite{CarParrinello} to compute the electronic ground state, \ac{BOMD} calculations, and \ac{WC}s. The Becke-Lee-Yang-Parr (BLYP) functional~\cite{lee1988Development,becke1988Densityfunctional} within the generalized gradient approximation (GGA) framework with Grimme's dispersion correction D2~\cite{grimme2006Semiempirical} was used for the exchange-correlation functional, and the Goedecker-Teter-Hutter (GTH) pseudopotentials~\cite{goedecker1996Separablea} was employed with the plane wave cutoff of $\SI{100}{\mathrm{Ry}}$. The BLYP functional has been used regularly in past \ac{AIMD} simulations of methanol~\cite{handgraaf2003Initio,handgraaf2004Densityfunctional,woods2005Influence,mcgrath2011Vapor, mcgrath2011Liquid, yadav2012First, wang2017Initio, jindal2020Geometry}. The energy convergence conditions were $\SI{1e-7}{\hartree}$ for the \ac{ML} training data and $\SI{1e-4}{\hartree}$ for the \ac{BOMD} calculations. Only $\Gamma$ point was used in the electron density integration. Lattice constants were determined from experimental densities~\cite{assessment2009CRC} for liquid simulations with $48$ molecules for methanol and $32$ for ethanol, while they were fixed at $\SI{20}{\angstrom}$ for gas phase calculations.

We employed the \ac{CMD} calculations with the GROMACS package~\cite{abraham2015GROMACS} to prepare \ac{ML} training structures and initial configurations for the \ac{AIMD} simulations. The general AMBER force field (GAFF2)~\cite{wang2004Development} and the AM1-BCC charge~\cite{jakalian2002Fast} were used. Topology files were generated by the Antechamber package~\cite{wang2006Automatic} and the ACPYPE package~\cite{sousadasilva2012ACPYPE}. Initial systems were made by randomly placing molecules using the Packmol package~\cite{martinez2009PACKMOL} with the density fixed to the experimental values. The simulations were performed in an NVT ensemble at $\SI{300}{\kelvin}$ using velocity rescaling. The MD time step was $\SI{1}{\femto\second}$. Ten thousand training data structures for each material were sampled every $\SI{1}{\pico\second}$ so that each structure was as uncorrelated as possible for both gas and liquid phases.

Calculated \ac{WC}s were assigned to each chemical bond by accommodating the closest WC to each \ac{BC}. Embedding and fitting \ac{ML} models, coded using the Pytorch library~\cite{paszke2019PyTorch}, contain three hidden layers with $50$ nodes. We chose the ReLU function~\cite{agarap2019Deep} as the activation function for all hidden layers. The hyperparameters were set to $M=20$ and $M'=6$ to balance computation time and accuracy. The cutoffs for the descriptors were set to $\SI{4}{\angstrom}$ for the inner cutoff and $\SI{6}{\angstrom}$ for the outer cutoff. The Adam stochastic gradient descent method~\cite{kingma2017Adam} was adopted for the optimization, where a mini-batch learning scheme with a batch size of $32$ was used. We independently trained the \ac{ML} models for methanol and ethanol, gas and liquid phases. The code is implemented in the MLWC (machine learning Wannier center) package~\cite{ToAmano}.

\ac{BOMD} simulations combined with \ac{ML} dipole moments were carried out to calculate dielectric properties. After initial configurations were generated from $\SI{10}{\nano\second}$ \ac{CMD} calculations, we performed a $\SI{5}{\pico\second}$ equilibration run by \ac{BOMD}, from which production runs of $\SI{10}{\pico\second}$ were carried out. All the \ac{BOMD} simulations were conducted in an NVT ensemble at $\SI{300}{\kelvin}$ with a Nos\'{e}-Hoover thermostat, with the integration time step of $\SI{0.25}{\femto\second}$. 

As the estimation of $\varepsilon^{\infty}$ was not within the scope of this work, we evaluated it from the square of the refractive index $n$ as $\varepsilon^{\infty}=n^2$. The experimental values~\cite{assessment2009CRC} of  $n=1.329$ for methanol and $n=1.361$ for ethanol at $\SI{298}{\kelvin}$ were used, noting that the temperature dependence of the refractive index is small.

\subsection{Model accuracy for gas phase}

\begin{figure}[htb]\centering
\captionsetup[subfigure]{labelformat=empty}
\centering
\begin{subcaptionblock}{\linewidth}
  \subcaption{}
  \adjustbox{right}{\includegraphics[width=\textwidth]{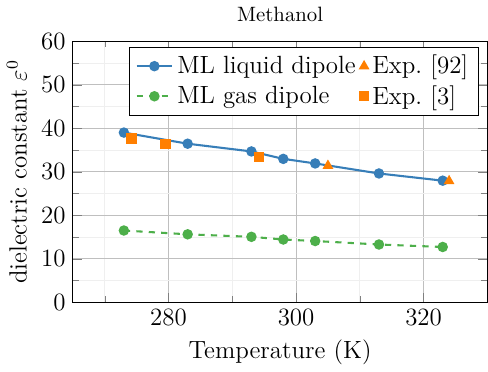}}
 \end{subcaptionblock}\hfill
\begin{subcaptionblock}{\linewidth}
 \subcaption{}
  \adjustbox{right}{\includegraphics[width=\textwidth]{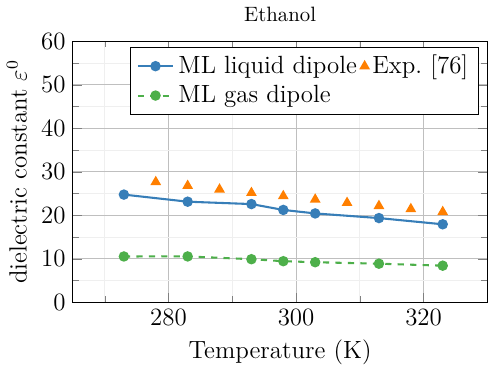}}
 \end{subcaptionblock}\hfill
\caption{Temperature dependence of the dielectric constant of methanol (top) and ethanol (bottom) calculated from liquid (blue) and gas (green) models using $\SI{50}{\nano\second}$ \ac{CMD} trajectories. The experimental values (orange) are taken from Refs~\cite{davidson1957DIELECTRIC,franck1978Dielectric,gregory2001Tables}.}
\label{fig:dielconst_met}
\end{figure} 

\begin{table*}[tb]
\centering
\caption{Calculated dielectric constant of liquid methanol at $\SI{298}{\kelvin}$ using various \ac{ML} models at $\SI{298}{\kelvin}$ with experimental data~\cite{gregory2001Tables}. $\mathrm{R}$ stands for alkyl chains, $\ce{CH3}$ for methanol and $\ce{CH3CO}$ for ethanol. The superscripts G and L represent the predicted values by the gas and liquid models.}
\input{dielconst.tex}
\label{table:dielconst}
\end{table*} 

\begin{figure*}[thb]  
\captionsetup[subfigure]{font={bf,large}, skip=1pt, margin=-0.7cm,justification=raggedright, singlelinecheck=false}
\centering
\begin{subcaptionblock}{0.48\linewidth}
\adjustbox{right}{\includegraphics[width=\textwidth]{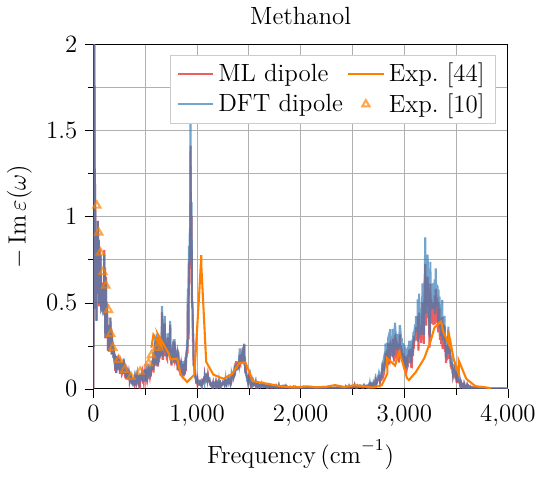}}
 \end{subcaptionblock}\hfill
 \begin{subcaptionblock}{0.48\linewidth}
\adjustbox{right}{\includegraphics[width=\textwidth]{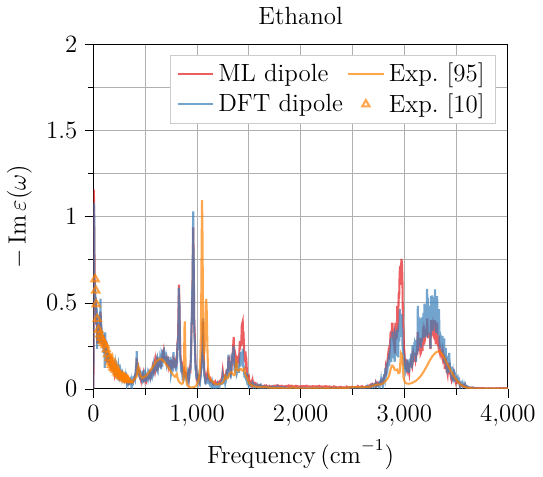}}
\end{subcaptionblock}
\caption{Calculated imaginary part of dielectric functions of methanol (left) and ethanol (right) accompanied by experimental values (orange)~\cite{wang2017Initio,sani2016Spectral,sarkar2017Broadband} at room temperature. All trajectories utilized $\SI{10}{\pico\second}$ \ac{BOMD} calculations and dipoles were calculated both from ML (red) and DFT (blue).}
\label{fig:dielec_met}
\end{figure*} 

\begin{figure}[thb]  
\centering
\adjustbox{right}{\includegraphics[width=\linewidth]{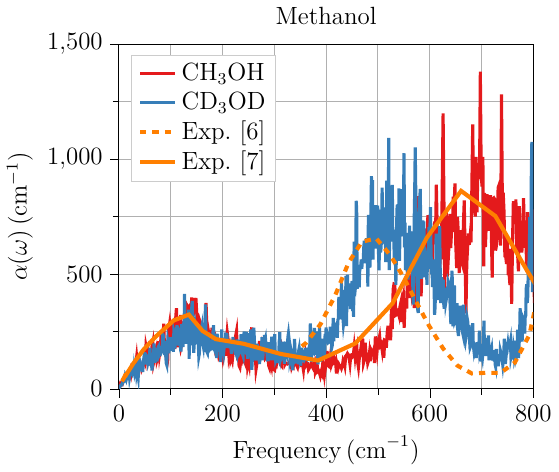}}
\caption{Calculated absorption $\alpha(\omega)$ of methanol (red) and deuterated methanol (blue), which were computed with \ac{BOMD} trajectories and \ac{ML} dipole models. The experimental spectra of methanol (orange)~\cite{bertie1993Infrared} and deuterated methanol (orange dashed)~\cite{bertie1994Infrared} are also shown.} 
\label{fig:deutron}
\end{figure}

\begin{figure}[tb]  \centering
\adjustbox{right}{\includegraphics[width=\linewidth]{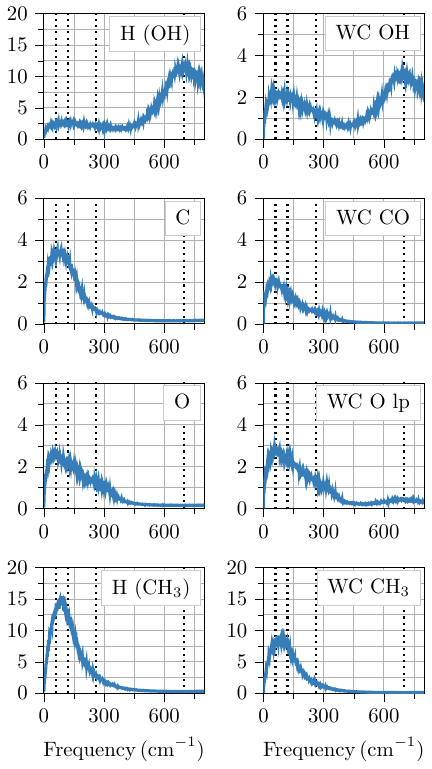}}
\caption{Analysis of the vibrational density of states (VDOS) of liquid methanol. Each atomic and \ac{WC} contributions are reported, where the hydroxyl hydrogen and the \ce{OH} bond \ac{WC} have a significant peak at $\SI{700}{\cm^{-1}}$. The vertical dotted lines represent experimentally observed peak positions at $60$, $120$, $260$, and $\SI{700}{\per\cm}$.}
 \label{fig:vdos}
\end{figure} 

\begin{figure}[thb]  \centering
\adjustbox{right}{\includegraphics[width=\linewidth]{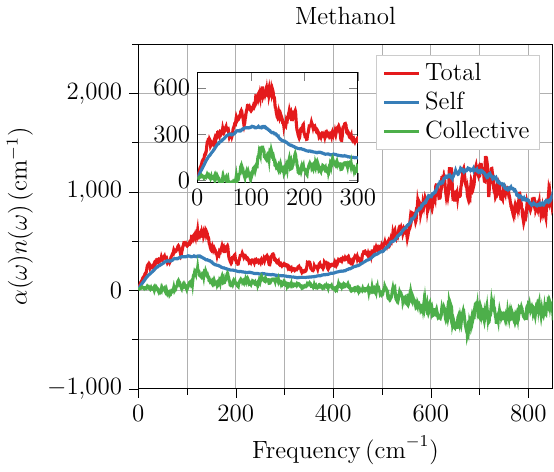}}
\caption{Calculated absorption $\alpha(\omega)n(\omega)$ of methanol (red), decomposed into self (blue) and collective (green) components according to \cref{180524_26Feb24}. We use the centered moving average method to smooth the spectra.}
\label{fig:cross_self}
\end{figure} 

We have four \ac{ML} dipole models for methanol, namely the \ce{CH} bond, CO bond, OH bond, and \ac{O-lp} models, whereas there is an additional \ce{CC} bond model for ethanol. We first discuss the results of the isolated (gas phase) methanol and ethanol. \Cref{fig:lr_gas_met} compares dipole moments calculated from first principles with those predicted by the \ac{ML} models, showing excellent agreement with each other. As there is no polarization owing to intermolecular interactions in isolated systems, \ac{ML} models only need to learn the effect of intramolecular deformations on the displacement of the \acp{WC}. We note that the inner cutoff of $\SI{4}{\angstrom}$ for the descriptor is sufficiently large that the descriptors include all atoms. \Cref{table:lr} summarizes the \ac{RMSE} of each model, defined as
\begin{align}
 \mathrm{RMSE} = \sqrt{\frac{1}{n}\sum_{i=1}^{n}\left|\vb*{\mu}^{\mathrm{p}}_{i}-\vb*{\mu}^{\mathrm{t}}_{i}\right|^2},
\end{align}
where $n$ is the number of data. All \ac{ML} models perform well with an accuracy of almost $\mathrm{RMSE}=\SI{0.03}{\mathrm{D}}$ or less, noting that the bond dipole moment of $\SI{1}{\mathrm{D}}$ corresponds to about $\SI{0.1}{\angstrom}$ of the displacement of \ac{WC}.

\subsection{Model accuracy for liquid phase}
We next discuss the results for liquid systems. Intermolecular interactions cause significant polarization in liquids, which makes prediction more difficult. \Cref{fig:lr_met} compares dipole moments calculated from first principles with those predicted by the ML models, showing good agreement with each other. \Cref{table:lr} summarizes the \ac{RMSE} of each model, with errors cured to less than $\SI{0.04}{\mathrm{D}}$. The \ac{O-lp} model is the least accurate for methanol and ethanol in common because the \acp{WC} of the \ac{O-lp} frequently move in liquids due to the electric field created by the surrounding molecules. Therefore, the prediction accuracy of the dipole moments of the system is sensitive to that of the \ac{O-lp}.

\Cref{fig:lr_met} also shows the prediction accuracy of the molecular dipole moments of liquid methanol and ethanol. For example, the dipole moments of methanol molecules are the sum of the corresponding bond dipole moments using \cref{Eq:bd} as 
\begin{align}
 \vb*{\mu}^{\mathrm{mol}} = \vb*{\mu}^{\mathrm{Olp}} + \vb*{\mu}^{\mathrm{OH}} + \vb*{\mu}^{\mathrm{CO}} + \vb*{\mu}_{1}^{\mathrm{CH}} + \vb*{\mu}_{2}^{\mathrm{CH}} + \vb*{\mu}_{3}^{\mathrm{CH}}.\label{043538_27Feb24} 
\end{align}
The \ac{RMSE} for the methanol molecular dipole moments is $\SI{0.09}{\mathrm{D}}$ and that for ethanol is $\SI{0.14}{\mathrm{D}}$, which are very accurate given that the average molecular dipole moments are about $\SI{2.5}{\mathrm{D}}$ for both materials. 

The largest differences between isolated and liquid systems appear in the \ac{O-lp} and OH models, which can be understood from the presence of hydrogen bonding in liquids. The \ac{O-lp} dipole moments, which averaged about $\SI{2.8}{\mathrm{D}}$ in the gas phase, increase to $\SI{3.7}{\mathrm{D}}$ in the liquid phase, and the dipole moments of the OH bonds, which averaged about $\SI{0.5}{\mathrm{D}}$ in the gas phase, take values ranging from $\SI{0}{\mathrm{D}}$ to $\SI{1}{\mathrm{D}}$. Additionally, the \ac{O-lp} dipole moments are most broadly distributed, indicating that the effect of the surrounding molecules can significantly change the positions of \acp{WC}. In contrast, the values of the \ce{CO}, \ce{CH}, and \ce{CC} dipole moments vary slightly from the gas phase, suggesting that the \acp{WC} of these bonds are insusceptible to neighbor molecules.

\subsection{The dipole moments}

We utilized the gas and liquid ML models to examine the effect of intermolecular interactions in liquids on dipole moments. \Cref{fig:dipole_gas_liquid} compares the molecular dipole moments of liquid molecular structures predicted from both gas (orange) and liquid (blue) models, where $1000$ molecular structures were randomly sampled from the $\SI{10}{\pico\second}$ \ac{BOMD} trajectory at $\SI{300}{\kelvin}$. The average molecular dipole moments from liquid models were approximately $\SI{1}{\mathrm{D}}$ larger than those from gas models in both methanol and ethanol, as in previous studies~\cite{handgraaf2003Initio,jorge2022Dipole}. Since the molecular structures are taken from simulated liquids, the difference entirely stems from the polarization of the \acp{WC} due to intermolecular interactions. \Cref{table:dipole} summarizes the average molecular dipole moments, where $\mu^{\mathrm{mol}}_{\mathrm{G}}$ and $\mu^{\mathrm{mol}}_{\mathrm{L}}$ stand for the dipole moments of the gas and liquid phases, respectively. Predicted values aligned well with experimental values~\cite{assessment2009CRC} in the gas phase. For the liquid phase of methanol, previous DFT simulations using the same functional reported $\SI{2.54}{\mathrm{D}}$~\cite{handgraaf2003Initio} or $\SI{2.68}{\mathrm{D}}$~\cite{sieffert2013Liquid}, which were consistent with our result of $\SI{2.69}{\mathrm{D}}$.

To examine which bond dipole moments have large difference between gas and liquid models, \cref{fig:dipole_gas_liquid_2} compares the bond dipole distributions of $\vb*{\mu}^{\mathrm{\ce{CH3CO}}}=\vb*{\mu}^{\mathrm{\ce{CH3}}}+\vb*{\mu}^{\mathrm{\ce{CO}}}$ (orange), $\vb*{\mu}^{\mathrm{hydroxy}}=\vb*{\mu}^{\mathrm{\ce{OH}}}+\vb*{\mu}^{\mathrm{\ce{O}lp}}$ (red), and $\vb*{\mu}^{\mathrm{mol}}$ (blue) calculated from both liquid and gas models using liquid methanol structures. We found that $\vb*{\mu}^{\mathrm{hydroxy}}$ is significantly enhanced in liquid models, while $\vb*{\mu}^{\mathrm{\ce{CH3CO}}}$ shows less variation. 

\Cref{fig:o_wc_rdf} shows the \ac{RDF} from \ce{O} atoms to \acp{WC} in liquid methanol. The three peaks correspond to the O \ac{lp}, \ce{OH} bond, and \ce{CO} bond, in ascending order of distance. Comparing the peak positions in the liquid phase with those in the gas phase (vertical dotted lines), we found that the \acp{WC} of O \ac{lp} move further away from the O atom, while the \acp{WC} of the \ce{OH} bond move closer to the \ce{O} atoms in the liquid phases, as shown in previous studies~\cite{silvestrelli1999Water,handgraaf2003Initio}, thus increasing $\vb*{\mu}^{\mathrm{hydroxy}}$.

\subsection{The dielectric constants}

\begin{figure*}[thb]  \captionsetup[subfigure]{font={bf,large}, skip=1pt, margin=-0.7cm,justification=raggedright, singlelinecheck=false}
\centering
\begin{subcaptionblock}{0.5\linewidth}
  \adjustbox{right}{\includegraphics[width=\textwidth]{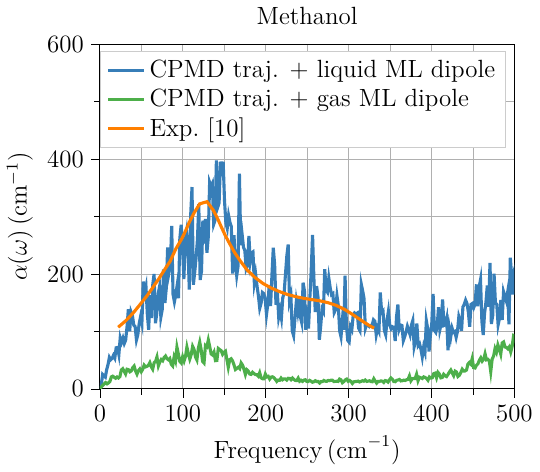}}
\end{subcaptionblock}\hfill
\begin{subcaptionblock}{0.5\linewidth}
  \adjustbox{right}{\includegraphics[width=\textwidth]{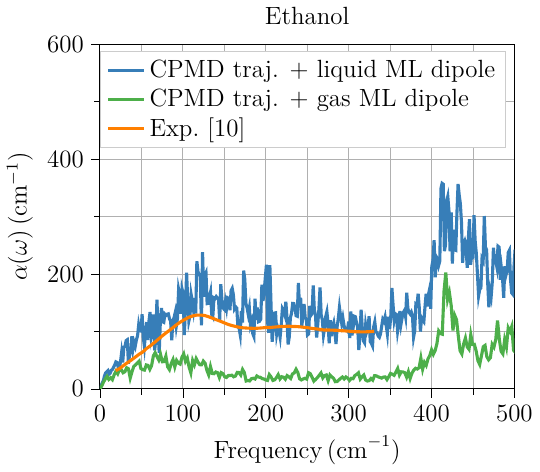}}
\end{subcaptionblock}
\caption{Calculated absorption $\alpha(\omega)$ of methanol (left) and ethanol (right) from \ac{BOMD} trajectories and \ac{ML} dipole moments. The liquid model calculations (blue) agree well with experimental values (orange) at room temperature from Sarker et al.~\cite{sarkar2017Broadband}, while the gas model calculations (green) underestimate them.}
\label{fig:alpha_met}
\end{figure*} 

\begin{figure}[tb]
\centering
  \includegraphics[width=0.6\linewidth]{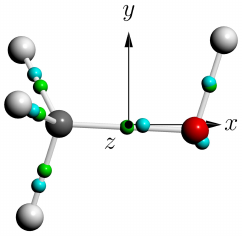}
  \caption{Principal inertia axes of methanol. We define $x$, $y$, and $z$ axes according to the decreasing order of their principal values.}
\label{Fig:inertia}
\end{figure} 

\begin{figure}[thb]  
\captionsetup[subfigure]{font={bf,large}, skip=1pt, margin=-0.7cm,justification=raggedright, singlelinecheck=false}
\centering
\adjustbox{right}{\includegraphics[width=\linewidth]{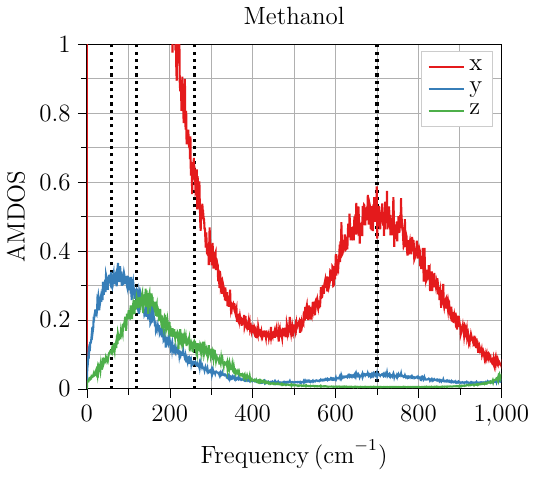}}
\adjustbox{right}{\includegraphics[width=\linewidth]{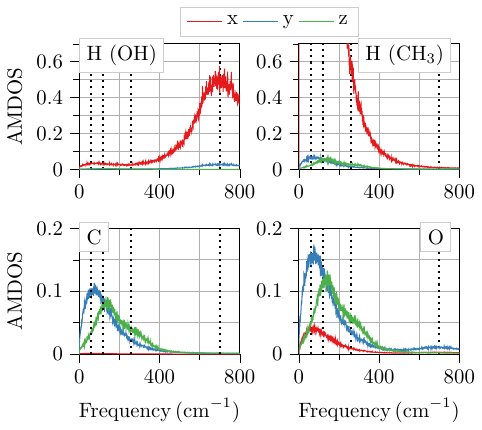}}
\caption{Fourier spectra of the angular velocity auto-correlations around three principal axes of $x$ (red), $y$ (blue), and $z$ (green) using \cref{021406_28Feb24}. The full spectrum is displayed at the top, followed by the decomposition in terms of the atomic contributions. The vertical dotted lines represent experimentally observed peak positions at $60$, $120$, $260$, and $\SI{700}{\per\cm}$. We use the centered moving average method to smooth the spectra.}
\label{fig:angular_momentum}
\end{figure} 

Since long trajectories are required to calculate dielectric constants~\cite{krishnamoorthy2021Dielectric}, we performed $\SI{50}{\nano\second}$ \ac{CMD} calculations and predicted the dipole moment every $\SI{1}{\pico\second}$ using $333$ and $500$ molecule systems for methanol and ethanol, respectively. The timescale is far beyond the DFT capacity. We used the experimental densities at each temperature~\cite{freyer1929SONIC} ranging from $\SI{273}{\kelvin}$ to $\SI{323}{\kelvin}$ to determine the lattice constants. We evaluated the static dielectric constants using \cref{171945_26Feb24}, which does not contain dynamical information and requires only sufficient structural sampling at equilibrium for convergence. We achieved rapid structure sampling through \ac{CMD} and accurate dipole moment calculations via \ac{ML} models, accounting for intermolecular interaction effects. 

\Cref{fig:dielconst_met} shows that predicted dielectric constants using liquid models (blue) agree well with experimental data (orange)~\cite{franck1978Dielectric,davidson1957DIELECTRIC,gregory2001Tables}, while calculated values from the gas models (green) underestimate the dielectric constants by more than $50\%$ for both methanol and ethanol, indicating the importance of \ac{WC} polarization on the dielectric constant. Some classical empirical point charges are known to underestimate the dielectric constant of methanol~\cite{shilov2015Molecular}, which may be attributed to a similar cause. Also, our models were able to be used over various temperatures and densities, even though they were trained on structures only at $\SI{300}{\kelvin}$. Ethanol might exhibit greater sensitivity to the choice of force field because of its more complex structure than methanol, which potentially contributed to the underestimation of our calculations.

\Cref{table:dielconst} summarizes the dielectric constants at $\SI{298}{\kelvin}$ calculated by the liquid models, gas models, and their combinations, where we divided the molecules into an alkyl chain and the hydroxy part (OH+O\ac{lp}) and evaluated each part using the gas and liquid models. For methanol, the combination of the gas model evaluation for alkyl chains and liquid model calculation of the hydroxy part yields a dielectric constant of $26.7$, which is close to the complete liquid calculation of $32.21$. Therefore, the hydroxy part explains most of the dielectric constant, while the polarization of alkyl chains in liquids also has a non-negligible contribution. The same is true for ethanol.

\subsection{The dielectric functions}

\Cref{fig:dielec_met} shows the imaginary part of the dielectric function calculated from DFT (blue) and ML (red) in the IR region up to $\SI{4000}{\cm^{-1}}$, alongside experimental values (orange) at room temperature~\cite{wang2017Initio,sarkar2017Broadband,doroshenko2012Infrared}. We averaged five independent \ac{BOMD} trajectories of $\SI{10}{\pico\second}$ to smooth the spectra, with the dipole moments evaluated using both DFT and ML models. Therefore, the difference between the ML and DFT calculations appears in the intensity of the spectra. Overall, \ac{ML} results reproduced DFT calculations for both methanol and ethanol. Additionally, our calculations agree well with the experimental data, although some peaks show a redshift, which is characteristic of the BLYP functional~\cite{wang2017Initio}. Peak assignments have been done experimentally~\cite{plyler1952infrared}: the $\SI{2900}{\cm^{-1}}$ and $\SI{3300}{\cm^{-1}}$ peaks are attributed to the stretching motion of \ce{CH} and \ce{OH}, respectively, while the $\SI{1300}{\cm^{-1}}$ peak is ascribed to the bending of \ce{OH}, and the peaks around $\SI{1000}{\cm^{-1}}$ are of methyl groups motions.

To study THz spectra below $\SI{1000}{\cm^{-1}}$, we prepared five independent $\SI{20}{\pico\second}$ \ac{BOMD} trajectories with \ac{ML} dipole moments. First, we focus on the libration peak at $\SI{700}{\cm^{-1}}$. \Cref{fig:deutron} shows that the absorption spectra of deuterated and normal methanol agree well with experimental values at $\SI{298}{\kelvin}$~\cite{bertie1993Infrared,bertie1994Infrared}. The $\SI{700}{\cm^{-1}}$ peak of normal methanol is shifted to about $\SI{500}{\cm^{-1}}$ in deuterated methanol, which verifies that the $\SI{700}{\cm^{-1}}$ peak is due to hydrogen motions. In addition, \cref{fig:vdos} shows the partial VDOS of the atoms and \acp{WC} using~\cref{063801_26Feb24}, from which we found that the hydroxyl hydrogens and \acp{WC} exhibit a strong peak at $\SI{700}{\cm^{-1}}$. Therefore, the $\SI{700}{\cm^{-1}}$ peak is due to the motion of hydrogens in the \ce{OH} bond, as pointed out in previous studies~\cite{skaf1993Wave,torii2023Intermolecular}.

\Cref{fig:cross_self} shows the decomposition of simulated $\alpha(\omega)n(\omega)$ into self (blue) and collective (green) components according to \cref{180524_26Feb24}. The $\SI{700}{\cm^{-1}}$ libration peak is mostly described by the self component, whereas the collective (intermolecular) and self (intramolecular) components have equal positive contributions up to $\SI{300}{\cm^{-1}}$. 

\Cref{fig:alpha_met} demonstrates that the calculated absorption coefficient $\alpha(\omega)$ up to $\SI{500}{\cm^{-1}}$ using liquid \ac{ML} models (blue) excellently agrees with experimental values (orange)~\cite{sarkar2017Broadband}. In contrast, the gas model calculations (green) considerably underestimate the experimental data. The $\SI{426}{\cm^{-1}}$ peak seen only in ethanol is due to the \ce{CCO} bending~\cite{plyler1952infrared}. Of the $60$, $120$, and $\SI{260}{\cm^{-1}}$ peaks of methanol noted in previous studies~\cite{sarkar2017Broadband,yomogida2010Dielectric}, the former two appear as one solid peak on the low-frequency side, while the highest peak is very broad. In addition, the gas model calculations partially reproduce the first peak and fail to reproduce the last broad peak. Therefore, internal molecular motions can partly explain the first two peaks, while the broad peak is entirely due to intermolecular interactions.

To analyze the intramolecular motions, we performed the principal axes analysis of the angular velocity ACF of methanol. As shown in \cref{Fig:inertia}, we define $x$, $y$, and $z$-axes according to the increasing order of its principal values. The $x$-axis is approximately parallel to the \ce{CO} bond, while the $y$-axis is almost in the \ce{COH} plane. The $z$-axis is an out-of-plane vector orthogonal to these. \Cref{fig:angular_momentum} shows that the angular velocity ACFs in principal axes have four significant features: the peaks at $60$, $120$, and $\SI{700}{\cm^{-1}}$ for $y$, $z$, and $x$ components, respectively, and the large values on the $x$-axis in the low-frequency region below $\SI{300}{\cm^{-1}}$. The highest $\SI{700}{\cm^{-1}}$ peak is due to the rotational motion of the hydroxyl \ce{H} atom around the $x$-axis, consistent with the conclusion in \cref{fig:deutron} and \cref{fig:vdos}. The $60$ and $\SI{120}{\cm^{-1}}$ peaks arise from the rotational motion of carbon, oxygen, and alkyl hydrogen around $y$ and $z$-axes, respectively. For the latter, the phenomenon that the $x$-axis ACF of alkyl hydrogens takes large values at low frequencies, which has not been reported in previous rigid molecule calculations~\cite{matsumoto1990Hydrogen,venables2000Structure}, has little effect on the dielectric spectra, as it does not generate dipole moments. 

In summary, the THz absorption spectrum of methanol is characterized by the libration peaks at $60$ and $\SI{120}{\cm^{-1}}$, which are partially described by intramolecular vibrations, the libration peak at $\SI{700}{\cm^{-1}}$, which is almost completely described by intramolecular vibrations, and the broad peak at $\SI{260}{\cm^{-1}}$, which is completely described by intermolecular interaction. Alkyl hydrogen vibrations are also observed, but they do not significantly affect the dielectric properties.

\section{Conclusion}\label{sec:theory}

We have constructed a versatile ML scheme to predict dipole moments for molecular liquids by attributing \acp{WC} to each chemical bond and creating \ac{ML} models that predict the \acp{WC} for each chemical bond. This scheme is applicable to any molecular system as long as the \acp{WC} can be assigned to chemical bonds. We applied the developed method to the primary alcohols methanol and ethanol, confirming its high accuracy in both the gas and liquid phases.

Using the developed method, we conducted the first-principles study on the dielectric properties of liquid methanol and ethanol. The dipole moment increases by approximately $\SI{1}{\mathrm{D}}$ in the liquid phase compared to the gas phase, due to the \acp{WC} polarization of \ac{O-lp} and \ce{OH} bonds rather than the alkyl chains. The \ac{ML} models, combined with \ac{CMD}, accurately reproduced the experimental values of the dielectric constants. In contrast, the gas model calculations underestimate the dielectric constant by more than $50\%$. These results highlight the importance of the \acp{WC} polarization due to local intermolecular interactions. 

The \ac{ML} models were sufficiently accurate to calculate the IR dielectric function across the entire frequency range. The calculated THz absorption spectra also agreed well with the experiments. The $\SI{700}{\cm^{-1}}$ peak was assigned to the libration motion of the hydroxyl hydrogen around the \ce{CO} axis, as pointed out in previous studies~\cite{venables2000Structure,torii2023Intermolecular}. The low-frequency peaks were more complex, with approximately half of their magnitude resulting from intermolecular interactions. In particular, the broad $\SI{260}{\cm^{-1}}$ peak was entirely due to intermolecular interactions. Intramolecular motions contribute to the out-of-plane ($y$-axis) librational peak at $\SI{60}{\cm^{-1}}$ and the in-plane ($z$-axis) librational peak at $\SI{120}{\cm^{-1}}$. These analyses provide new insights into the origin of the THz spectrum of methanol and demonstrate the high accuracy of our \ac{ML} models. We expect that the presented approach will be valuable for predicting the dielectric properties of a wide range of materials.

\begin{acknowledgments}
This research was funded by a JST-Mirai Program Grant Number JPMJMI20A1, a MEXT Quantum Leap Flagship Program (MEXT Q-LEAP) grant number JPMXS0118067246, Japan, and JSR Corporation via JSR-UTokyo Collaboration Hub, CURIE. The computations in this study have been conducted using computational resources of the supercomputer Fugaku provided by the RIKEN Center for Computational Science (ProjectID: hp220331, hp230124) and the facility of the Supercomputer Center, the Institute for Solid State Physics, the University of Tokyo.
\end{acknowledgments}

\appendix

\section{Linear response theory}\label{appendix:A}

To consider how to calculate the dielectric function in MD simulations, we start with the linear response theory of polarization $\bar{\vb*{P}}(t)$ to an electric field $\vb*{E}(t)$,  \begin{align}
\bar{P}_{\alpha}(t)=\frac{1}{V} G^{R}(\hat{M}_{\alpha},\vb*{\hat{M}},\omega)\cdot \vb{E}(t),\label{165800_25Jan24}
\end{align}
where $V$ is the system volume, and $\vb*{\hat{M}}$ is the dipole moment operator. $\alpha$ and $\beta$ are the Cartesian indices. The retarded Green's function is given by \begin{align}
&G^R(\hat{M}_{\alpha},\hat{M}_{\beta},\omega) = \int_{0}^{\infty}e^{-i\omega t} G^R(\hat{M}_{\alpha}(t),\hat{M}_{\beta}(0)) \dd t \\
&G^R(\hat{M}_{\alpha}(t),\hat{M}_{\beta}(0))  = \frac{\theta(t)}{i\hbar}\expval{\left[\hat{M}_{\alpha}(t),\hat{M}_{\beta}(0)\right]}.\label{135251_29Jan24}
\end{align}
$\expval{}$ denotes the canonical ensemble average. Using the fact that the polarization and the electric field are connected by the dielectric susceptibility $\chi$ as $P_{\alpha}=\chi_{\alpha\beta}E_{\beta}$, and that the dielectric function  in the IR region is the sum of the atomic contribution $\chi$ and the electron contribution $\varepsilon^{\infty}$ as $\varepsilon_{\alpha\beta}=\varepsilon^{\infty}_{\alpha\beta}+\chi_{\alpha\beta}$, we write the dielectric function as
\begin{align}
 \varepsilon_{\alpha\beta}(\omega)=\varepsilon^{\infty}_{\alpha\beta}+\frac{1}{V} G^{R}(\hat{M}_{\alpha},\hat{M}_{\beta},\omega).
\end{align}
To utilize \cref{135251_29Jan24} in the MD calculations, we must get the classical limit of the Green's function, where the dipole moment is just a number. We used the famous harmonic approximation~\cite{ramirez2004Quantum,iftimie2005Decomposing,kubo1957StatisticalMechanical}, which replace the canonical correlation function~\cite{kubo1957StatisticalMechanical} with classical correlation function. The canonical correlation of two operators $A$ and $B$ is defined as
\begin{align}
 \expval{\hat{A};\hat{B}}_{\mathrm{can}} = \frac{1}{\beta}\int_{0}^{\beta}\dd\lambda \expval{e^{\lambda H_0}\hat{A}e^{-\lambda H_0}\hat{B}}\dd \lambda,
\end{align}
where $H_0$ is the unperturbed Hamiltonian and $\beta=1/k_{\mathrm{B}} T$ is the inverse temperature. To relate the canonical correlation with the retarded Green's function, we use the following equation
\begin{align}
 \left[\hat{A},e^{-\beta H_0}\right] &= e^{-\beta H_0}\int_{0}^{\beta} \dd \lambda e^{\lambda H_0}\left[H_0,\hat{A}\right]e^{-\lambda H_0} \\
 &= -i\hbar e^{-\beta H_0}\int_{0}^{\beta} \dd \lambda \dot{\hat{A}}(-i\hbar\lambda),\label{165007_25Jan24}
\end{align}
where $\hat{A}(t)=e^{-H_0t/i\hbar}\hat{A}(0)e^{H_0t/i\hbar}$ is the time-dependent operator, and the Heisenberg equation is used for the second line. The retarded Green's function can be written as
\begin{align}
 G^{R}(\hat{A}(t),\hat{B}) &= \frac{1}{i\hbar}\expval{\left[\hat{A}(t),\hat{B}(0)\right]} \\
 &= \frac{1}{i\hbar} \Tr \rho \left[\hat{A}(t),\hat{B}(0)\right] \\
 &= \frac{1}{i\hbar} \Tr \left[\rho, \hat{A}(t)\right] \hat{B}(0)\label{165013_25Jan24}
\end{align}
where $\rho = e^{-\beta H_0}/Z$ is the density matrix, and we used the cyclicity of the trace in the second line. Substituting \cref{165007_25Jan24} into \cref{165013_25Jan24}, we obtain the relation between the retarded Green's function and the canonical correlation function as
\begin{align}
 G^{R}(\hat{A}(t),\hat{B}) &= \Tr \rho \int_{0}^{\beta} \dd \lambda \hat{B}(-i\hbar\lambda) \dot{\hat{A}}(t) \\
 &= \beta \expval{\hat{B}; \dot{\hat{A}}(t)}_{\mathrm{can}} \\
 &= \beta \dv{t}\expval{\hat{B}; \hat{A}(t)}_{\mathrm{can}}.
\end{align}
In the harmonic approximation, we replace the canonical correlation function with the classical correlation function
\begin{align}
 C(t) = \expval{A(t)B(0)}.
\end{align}
As a result, the dielectric function becomes
\begin{align}
 \varepsilon_{\alpha\beta}(\omega)= \varepsilon^{\infty}_{\alpha\beta}-\frac{\beta}{V}\int_{0}^{\infty}\dv{\expval{M_{\alpha}(t)M_{\beta}(0)}}{t} e^{-i\omega t}\dd t.\label{132613_5Feb24}
\end{align}
The ensemble average can be calculated from the time average of the MD calculations with the assumption of ergodicity. The mean value of dipole moments is assumed to be zero. If the dipole has a nonzero mean value, we simply subtract the mean value from the total dipole as
\begin{align}
 \vb{\bar{M}} = \vb{M}-\expval{\vb{M}}.
\end{align}
The dielectric function of an isotropic and homogeneous fluid can be calculated by averaging diagonal components of \cref{132613_5Feb24} as
\begin{align}
 \varepsilon(\omega) &= \frac{\varepsilon_{11}(\omega)+\varepsilon_{22}(\omega)+\varepsilon_{33}(\omega)}{3} \\
 &=\varepsilon^{\infty}-\frac{1}{3k_{\mathrm{B}}TV}\int_{0}^{\infty}\dv{\expval{\vb{M}(t)\vdot\vb{M}(0)}}{t} e^{-i\omega t}\dd t.\label{151614_25Nov23}
\end{align}

\bibliographystyle{apsrev4-2}
%
%
\end{document}

%% file: lr.tex
 \renewcommand{\arraystretch}{1.5}
{\tabcolsep = 0.15cm
  \begin{tabular}{lccccccc}
   \hline\hline
   & \multirow{2}{*}{RMSE [D]} & \multicolumn{2}{c}{gas} && \multicolumn{2}{c}{liquid} & \\ \cline{3-4}\cline{6-7}
   &    & methanol & ethanol && methanol & ethanol& \\ \hline
   & CC & -       & $0.035$  && -       & $0.040$ & \\ 
   & CH & $0.004$ & $0.011$  && $0.030$ & $0.029$ & \\ 
   & CO & $0.011$ & $0.030$  && $0.023$ & $0.029$ & \\ 
   & OH & $0.010$ & $0.008$  && $0.022$ & $0.024$ & \\ 
   & O  & $0.023$ & $0.020$  && $0.040$ & $0.040$ & \\
    \hline\hline
  \end{tabular}
}
\renewcommand{\arraystretch}{1.0}

%% file: dipole.tex
 \renewcommand{\arraystretch}{1.5}
{\tabcolsep = 0.15cm
  \begin{tabular}{lccccccccccc}
   \hline\hline
    & \multicolumn{3}{c}{\ac{BOMD}} && \multicolumn{1}{c}{\ac{CMD}} && \multicolumn{4}{c}{Experiment} & \\ \cline{2-4}\cline{6-6}\cline{8-11}
    & $\mu^{\mathrm{mol}}_{\mathrm{G}}$ & $\mu^{\mathrm{mol}}_{\mathrm{L}}$ & $\Delta \mu^{\mathrm{mol}}$ &&  $\varepsilon^{0}$ && $\mu^{\mathrm{mol}}_{\mathrm{G}}$ & $\varepsilon^{0}$ & $n$ & $\rho$ & \\
   \hline 
   Methanol& $1.72$ & $2.69$ & $0.97$ &&  $33.0$ && $1.70$~\cite{assessment2009CRC} & $32.66$~\cite{gregory2001Tables} & $1.329$ & $0.7863$~\cite{hales1976Liquid} & \\ 
   Ethanol & $1.78$ & $2.71$ & $0.93$ &&  $21.2$ && $1.69$~\cite{assessment2009CRC} & $24.43$~\cite{gregory2001Tables} & $1.361$ & $0.7849$~\cite{hales1976Liquid} & \\ 
    \hline\hline
  \end{tabular}
}
 \renewcommand{\arraystretch}{1.0}

%% file: dielconst.tex
 \renewcommand{\arraystretch}{1.5}
{\tabcolsep = 0.15cm
  \begin{tabular}{lccccc}
   \hline\hline
             & gas  & $\mathrm{R}^{\mathrm{L}}+\mathrm{OH}^{\mathrm{G}}+\mathrm{Olp}^{\mathrm{G}}$ & $\mathrm{R}^{\mathrm{G}}+\mathrm{OH}^{\mathrm{L}}+\mathrm{Olp}^{\mathrm{L}}$ & liquid    & Exp.~\cite{gregory2001Tables} \\ \hline
    Methanol & $14.1$ & $18.1$ &  $26.7$ & $33.0$ & $32.66$ \\
    Ethanol  & $8.59$ & $10.8$ &  $18.3$ & $21.2$ & $24.43$ \\ 
    \hline\hline  
  \end{tabular}
}
\renewcommand{\arraystretch}{1.0}